\documentstyle[epsf]{article}

\begin{document}

\begin{center}

{\sf To be published in Physica A (2001): Proceedings of the NATO ARW} 
\underline{\sf on Application of Physics in Economic Modelling, Prague, Feb. 8-10, 2001}

\vspace*{1cm}

{\Large\bf Modeling electricity loads in California:\\ a continuous-time approach}

\vspace*{.6cm}

{\large Rafa{\l} Weron\footnote{Corresponding author. E-mail address: rweron@im.pwr.wroc.pl}}

\vspace*{.2cm}

{\small
Hugo Steinhaus Center for Stochastic Methods,\\
Wroc{\l}aw University of Technology, 50-370 Wroc{\l}aw, Poland}

\vspace*{.6cm}

{\large B. Koz{\l}owska, J. Nowicka-Zagrajek\footnote{Research partially supported by KBN Grant no. 8 T10B 034 17.}}

\vspace*{.2cm}

{\small
Institute of Mathematics,\\
Wroc{\l}aw University of Technology, 50-370 Wroc{\l}aw, Poland}

\end{center}

\vspace*{.2cm}

\begin{abstract}
In this paper we address the issue of modeling electricity loads and prices with
diffusion processes. More specifically, we study models which belong to the class 
of generalized Ornstein-Uhlenbeck processes. 
After comparing properties of simulated paths with those of deseasonalized data 
from the California power market and performing out-of-sample forecasts we conclude
that, despite certain advantages, the analyzed continuous-time processes are
not adequate models of electricity load and price dynamics.
\end{abstract}

\vspace*{.2cm}



\section{Introduction}

The last decade has witnessed radical changes in the structure of electricity markets
world-wide. Prior to the 1980s it was argued convincingly that the electricity industry 
was a natural monopoly and that strong vertical integration was an obvious and
efficient model for the power sector. In the 1990s, technological advances 
suggested that it was possible to operate power generation and retail supply as
competitive market segments \cite{ICC98,masson99}.

The changes that are taking place and the growing complexity of today's energy markets 
introduce the need for sophisticated tools for the analysis of market structures 
and modeling of electricity load and price dynamics \cite{kaminski99,bll99}.
However, we have to bear in mind that electricity markets are not anywhere near as
straightforward as financial or even other commodity markets. Demand and supply are 
balanced on a knife-edge because electric power cannot be economically stored, end 
user demand is largely weather dependent, and the reliability of the grid is paramount.

Recently it has been observed that, contrary to most financial assets \cite{bp97,ww98,ms99}, 
electricity price processes are mean-reverting \cite{kaminski99,pilipovic98,weron00,wp00}. 
In the next Sections we investigate whether electricity prices and loads in the California 
power market can be modeled by generalized Ornstein-Uhlenbeck processes, a special class 
of mean-reverting diffusion processes.

\section{Preparation of the data}

The analyzed database was provided by the University of California Energy Institute (UCEI)
\cite{UCEI}. Among others it contains market clearing prices from the California Power 
Exchange (CalPX) and system-wide loads supplied by California's Independent (Transmission) 
System Operator (ISO). At first we looked at CalPX clearing prices -- a time series 
containing system prices of electricity for every hour since April 1st, 1998, 0:00 until 
December 31st, 2000, 24:00. Because the series included a very strong daily cycle we  
created a 1006 days long sequence of average daily prices (as in \cite{weron00}), see 
Fig. 1. 

\begin{figure}[tbp]
\centerline{\epsfxsize=11cm \epsfbox{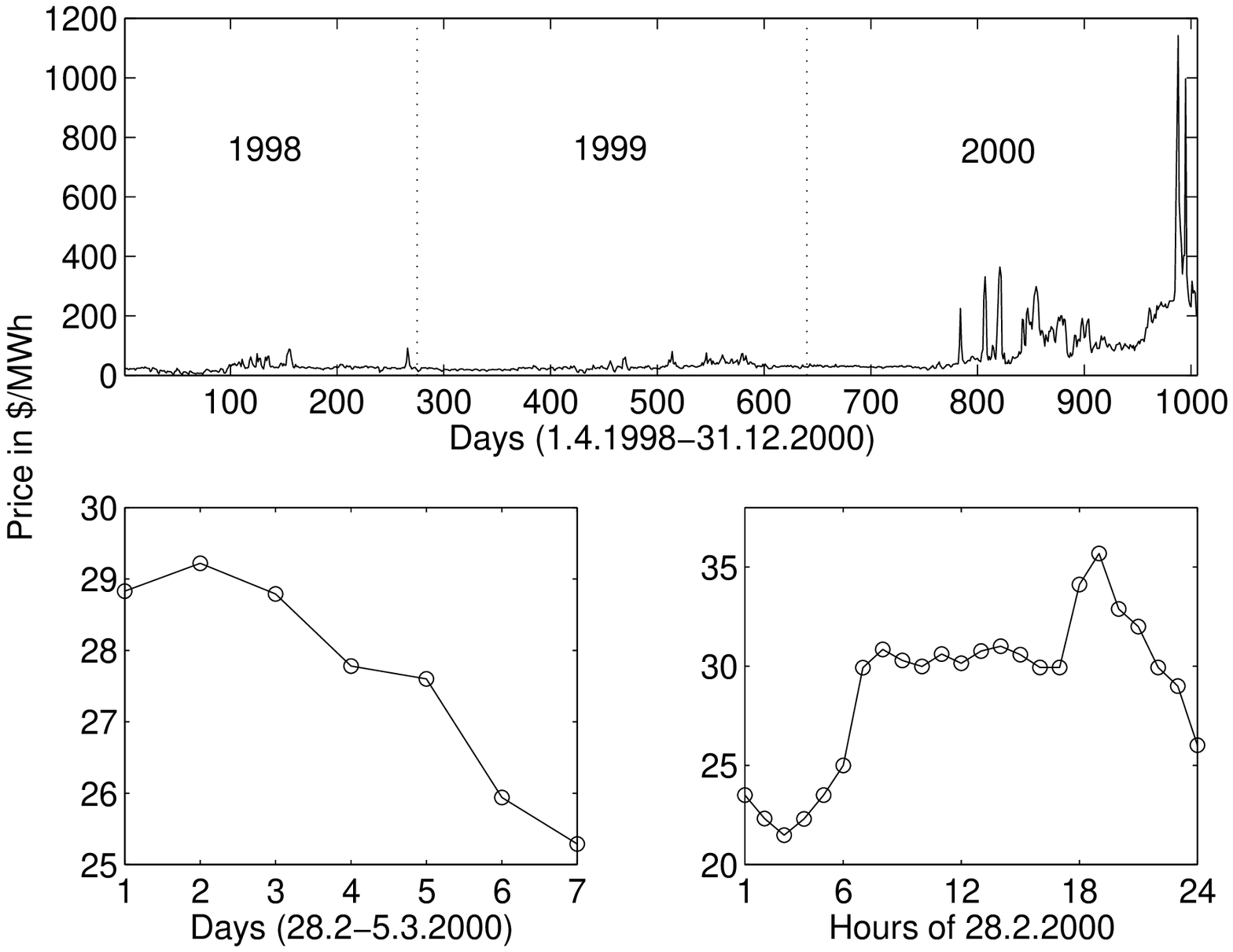}}
\caption{CalPX market daily average clearing prices since April 1st, 1998 until December 31st, 
2000 ({\it top panel}). A typical week in late winter -- CalPX market daily average 
clearing prices since February 28th, 2000 until March 5th, 2000 ({\it bottom left panel}).
The daily cycle -- low prices at night, high in the evening -- CalPX market hourly 
clearing prices on February 28th, 2000 ({\it bottom right panel}).
}
\end{figure}

\begin{figure}[tbp]
\centerline{\epsfxsize=11cm \epsfbox{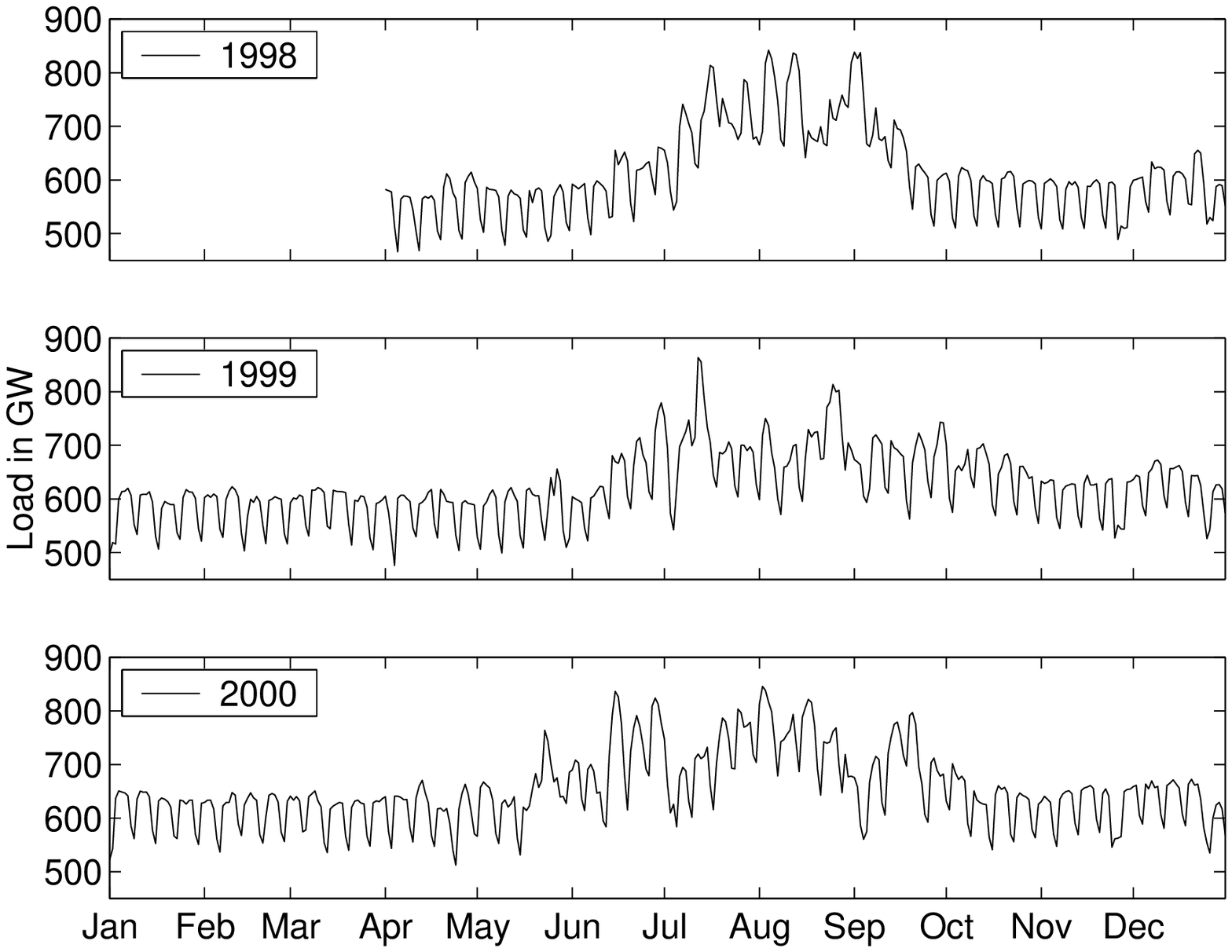}}
\caption{California power market daily system-wide load since April 1st, 1998 until 
December 31st, 2000. The annual and weekly seasonality are clearly visible.
}
\end{figure}

The price trajectory suggests that the process does not exhibit a regular annual cycle. 
Indeed, since June 2000, California's electricity market has produced extremely high 
prices and threats of supply shortages. The difficulties that have appeared are 
intrinsic to the design of the market, in which demand exhibits virtually no price
responsiveness and supply faces strict production constraints \cite{borenstein01}.
It is evident that without taking into consideration regulatory issues, modeling 
electricity prices in the "unstable" California power market is an almost impossible 
task. Thus, instead of forecasting electricity prices, we tried to tackle the "simpler" 
problem of modeling system loads. 

Like for electricity prices, the UCEI database contains information about the system-wide
load for every hour of the period April 1st, 1998 -- December 31st, 2000. 
Due to a very strong daily cycle we have created a 1006 days long sequence of daily loads,
which is plotted in Fig. 2. Apart from the daily cycle, the time series exhibits weekly 
and annual seasonality. 
Due to the fact that common trend and seasonality removal techniques do not work well 
when the time series is only a few (and not complete, in our case ca. 2.8 annual cycles) 
cycles long, we restricted the analysis only to two full years of data, i.e. to the period 
January 1st, 1999 -- December 31st, 2000, and applied a new seasonality reduction technique.

The seasonality can be easily observed in the frequency domain by plotting the periodogram, 
which is a sample analogue of the spectral density. 
For a vector of observations $\{x_1,...,x_n\}$ the periodogram is defined as
$I_n(\omega_k)=\frac{1}{n} \left|\sum_{t=1}^{n} x_t \exp\{-2\pi i (t-1) \omega_k\} \right|^2$,
where $\omega_k = k/n$, $k=1,...,[n/2]$ and $[x]$ denotes the largest integer less then or 
equal to $x$. Observe that $I_n$ is the squared absolute value of the Fourier transform. 
In order to use fast algorithms for the Fourier transform we restricted ourselves to 
vectors of even length, i.e. $n=2m$.
In Figure 3 we plotted the periodogram for the system-wide load before and after removal 
of the weekly and annual cycles. The periodogram shows well-defined peaks at frequencies 
corresponding to cycles with period 7 and 365 days. The smaller peaks close to $\omega_k=0.3$ 
and 0.4 indicate periods of 3.5 and 2.33 days, respectively. Both peaks are the so called 
harmonics (multiples of the 7-day period frequency) and indicate that the data exhibits 
a 7-day period but is not sinusoidal.
The weekly period was also observed in lagged autocorrelation plots \cite{weron00}. 
These cycles have to be removed before further analysis is carried out, since they may 
influence predictions to a great extent.

\begin{figure}[tbp]
\centerline{\epsfxsize=11cm \epsfbox{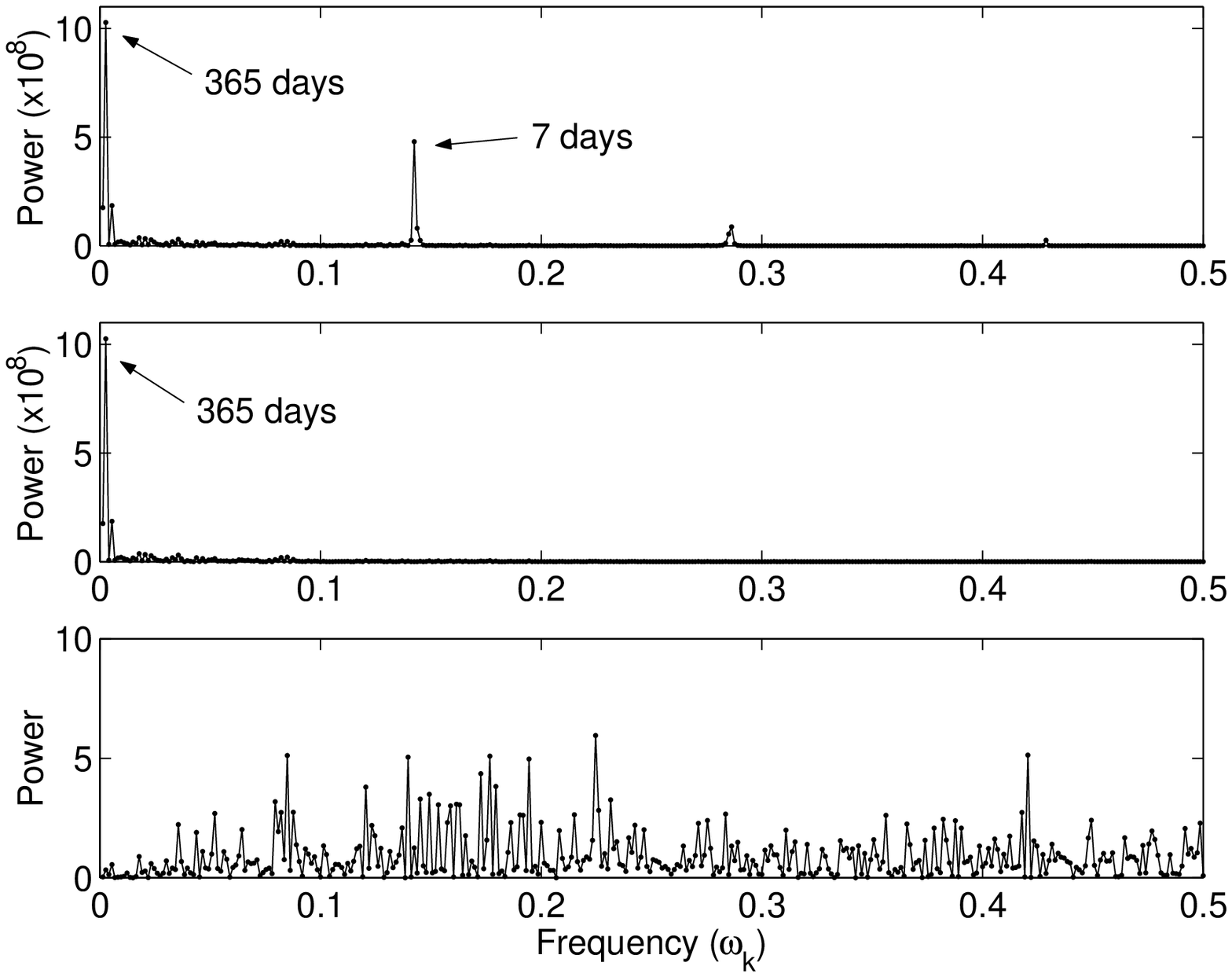}}
\caption{Periodogram of the California power market daily system-wide load since January 1st, 
1999 until December 31st, 2000 ({\it top panel}). The annual and weekly frequencies are clearly
visible. 
Periodogram of the load after removal of the weekly cycle ({\it middle panel}) and
of the load returns after removal of the weekly and annual cycles ({\it bottom panel}). 
In the last plot no dominating frequency can be observed.
}

\centerline{\epsfxsize=11cm \epsfbox{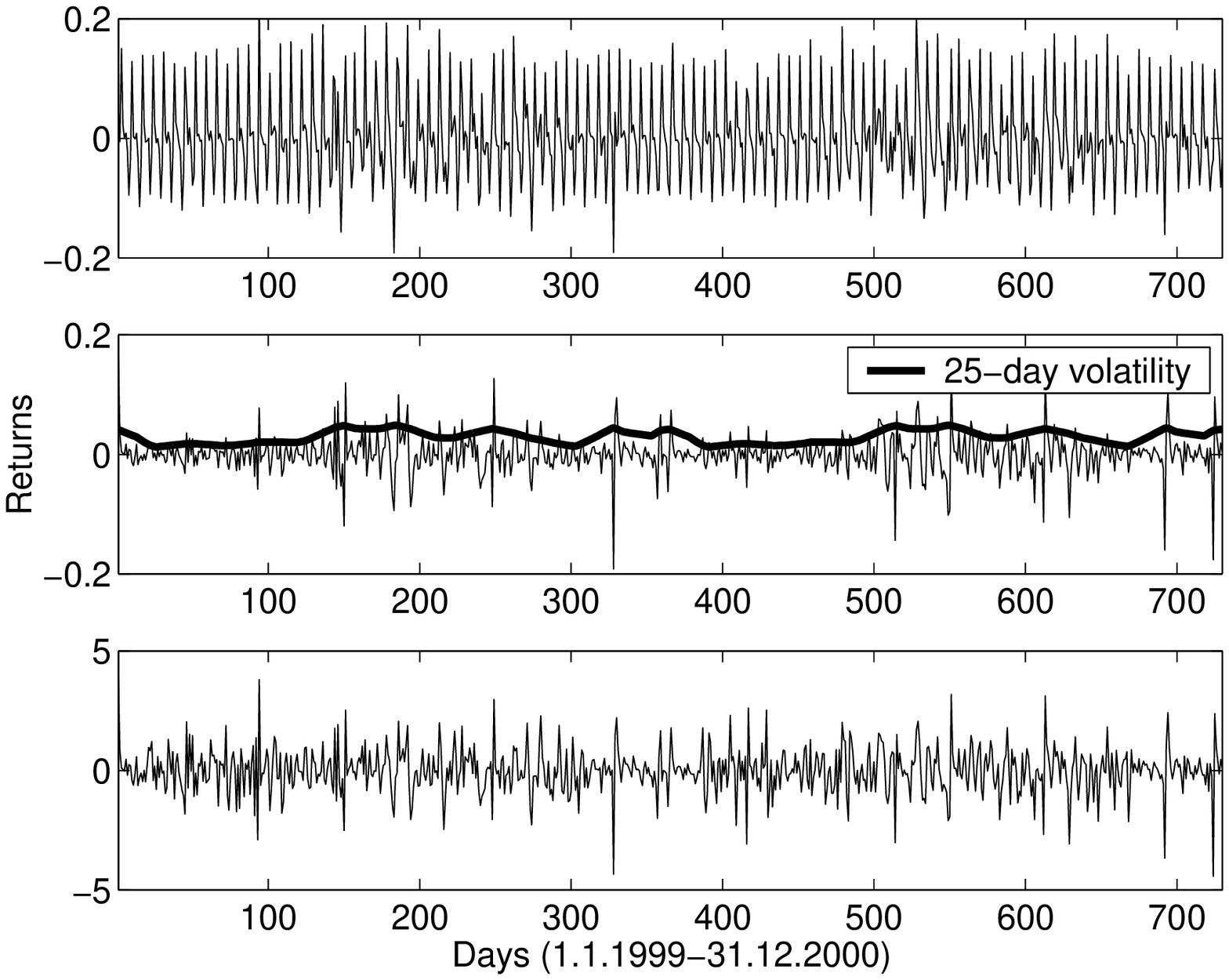}}
\caption{California power market daily system-wide load returns since January 1st, 1999 until 
December 31st, 2000 ({\it top panel}). 
Load returns after the removal of the weekly cycle and the 25-day rolling volatility 
({\it middle panel}). 
Load returns after removal of the weekly and annual cycles ({\it bottom panel}). 
}
\end{figure}

To remove the weekly cycle we used the moving average technique \cite{bd96}.
For the vector of daily loads $\{x_1,...,x_{730}\}$ the trend was first estimated by 
applying a moving average filter specially chosen to eliminate the weekly component 
and to dampen the noise: $\hat{m}_t=\frac17({x_{t-3}+...+x_{t+3}})$, where $t=4, ..., 727$. 
Next, we estimated the seasonal component. For each $k=1,...,7$, the average $w_k$ of the 
deviations $\{(x_{k+7j}-\hat{m}_{k+7j}), 3<k+7j\leq 727\}$ was computed. 
Since these average deviations do not necessarily sum to zero, we estimated the
seasonal component $s_k$ as $\hat{s}_k = w_k-\frac17\sum_{i=1}^{7}w_i$,
where $k=1,...,7$ and $\hat{s}_k=\hat{s}_{k-7}$ for $k>7$. 
The deseasonalized (with respect to the 7-day cycle) data was then defined as
$d_t=x_t-\hat{s}_t$ for $t=1,...,730$.
Finally we removed the trend from the deseasonalized data $\{d_t\}$ by 
taking logarithmic returns, see the middle panel of Fig. 4. 

After removing weekly seasonality we were left with the annual cycle. Unfortunately,
because of the short length of the time series (only two years), the method applied to 
the 7-day cycle could not be used to remove the annual cycle. To overcome this we introduced
a new method which consists of the following: (i) calculate a 25-day rolling volatility
\cite{kaminski97} for the whole vector; (ii) calculate the average volatility for one year
(i.e. in our case $v_t=(v_t^{1999}+v_t^{2000})/2$); (iii) smooth the volatility by taking 
a 25-day moving average of $v_t$; (iv) finally, rescale the returns by dividing them by 
the smoothed annual volatility. 
The obtained time series (see the bottom panel of Fig. 4) showed no apparent trend and
seasonality (see the bottom panel of Fig. 3). Therefore we treated it as a stationary process. 
In the next Section we fit the deseasonalized load returns by a generalized Ornstein-Uhlenbeck 
type model.

\section{Modeling with generalized Ornstein-Uhlenbeck type processes}

The deseasonalized data sets were modeled by mean-reverting continuous-type processes 
of the form (generalized Ornstein-Uhlenbeck processes):
\begin{equation}\label{gOU}
dX_t = \beta(m - X_t)dt + \rho X_t^{\gamma}dB_t.
\end{equation}
Unfortunately, since we were unable to remove the annual cycle from the system loads
themselves we had to restrict our analysis to models with $\gamma=0$ (we estimated  
$\gamma\in (0.3,0.8)$, but for fractional $\gamma$ the process has to be strictly positive
and evidently returns do not comply with this restriction). Thus we were left with 
the so-called Vasicek model \cite{vasicek77}. 

\begin{figure}[tbp]
\centerline{\epsfxsize=11cm \epsfbox{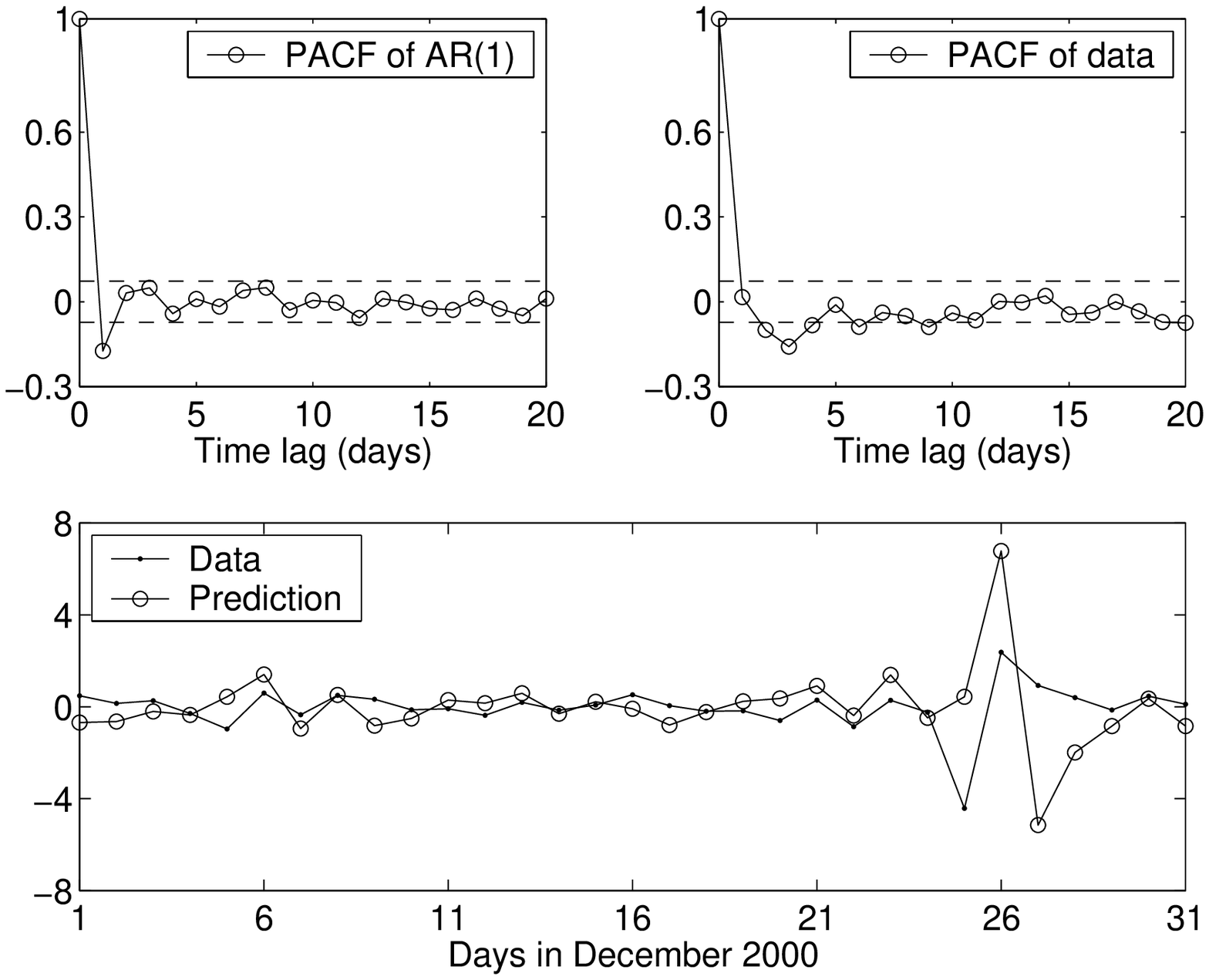}}
\caption{Partial autocorrelation function (PACF) of an AR(1) process ({\it top left panel})
and the deseasonalized system load logarithmic returns ({\it top right panel}).
Deseasonalized system load logarithmic returns in December 2000 and the Vasicek prediction 
({\it bottom panel}).
}
\end{figure}

We can calibrate the Vasicek model via ordinary linear regression: 
\begin{equation}
X_t = E(X_t) + \varepsilon_{t} = X_{t-1}e^{-\beta} + m(1-e^{-\beta}) + \varepsilon_t,
\end{equation}
where $\varepsilon_t \sim N(0,\rho_{\varepsilon})$ and $\rho_{\varepsilon}$ is 
the standard deviation from the regression.
Observe that the above implies that the Vasicek model is a continuous version of an AR(1)
process. This is the main reason why it performs poorly for our data sets. The deseasonalized
system loads may be an AR (Auto Regressive) process, however, of an order greater then 1 
(see the PACF plots in Fig. 5, which can be used as an estimate of the AR order \cite{bd96}). 
It is worth noting that other diffusions of the form (\ref{gOU}) also have a very short AR 
dependence structure, which would probably result in poor prediction of electricity prices
or loads.

For comparison, in the bottom panel of Fig. 5 we plotted actual deseasonalized load returns 
in December 2000 and the Vasicek prediction. The prediction is a one day forecast with 
model parameters estimated from the last 365 daily returns. Unfortunately the fit is far
from being perfect. The largest differences occur during Christmas (December 24th--26th), 
but this can be improved by incorporating a holiday structure into the model. 
However, the prediction for the first 23 days in December is still much worse than 
the prediction obtained from a simple ARMA(3,3) model \cite{nw01}, i.e. the mean absolute 
deviation from the true values is 0.565 compared to 0.355 for the ARMA forecast. 

Even though continuous-time models have certain advantages (like analytic tractability, 
a developed theory of pricing derivatives, etc. \cite{ww98,wilmott00}) over discrete models, 
further research will be in the direction of discrete time series models which offer
a much better fit to market data \cite{nw01}.

\end{document}